\documentclass[nofootinbib, superscriptaddress, twocolumn, 10pt]{revtex4-1}

\usepackage{url}
\usepackage{xcolor}
\usepackage{amsmath}
\usepackage{amssymb}
\usepackage{amsfonts}
\usepackage{hyperref}
\usepackage{booktabs}
\usepackage[english]{babel}
\usepackage{graphicx}
\usepackage[export]{adjustbox}
\graphicspath{{figs/}}

\usepackage{mathtools}
\DeclarePairedDelimiter\ceil{\lceil}{\rceil}

\usepackage{balance}

\usepackage{newtxmath}
\usepackage{mleftright}
\usepackage{dcolumn}
\usepackage{bm}

\newcommand{\RomanNumeralCaps}[1]
    {\MakeUppercase{\romannumeral #1}}

\newcommand{\si}[1]{\textcolor{blue}{\textit{SI Appendix}, #1}}

\begin{document}

\title{Community Structure Affects the Speed of Information Diffusion}
\title{Network Modularity Determines the Speed of Information Diffusion}
\title{Network Modularity Controls the Speed of Information Diffusion}

\author{Hao Peng}
\affiliation{School of Information, University of Michigan}
\author{Azadeh Nematzadeh}
\affiliation{S\&P Global}
\author{Daniel M. Romero}
\affiliation{School of Information, University of Michigan}
\author{Emilio Ferrara}
\thanks{Correspondence should be addressed to E.F.\\emiliofe@usc.edu}
\affiliation{Information Sciences Institute, University of Southern California}

\date{\today}
             
\begin{abstract}

The rapid diffusion of information and the adoption of social behaviors are of critical importance in situations as diverse as collective actions, pandemic prevention, or advertising and marketing. Although the dynamics of large cascades have been extensively studied in various contexts, few have systematically examined the impact of network topology on the efficiency of information diffusion. Here, by employing the linear threshold model on networks with communities, we demonstrate that a prominent network feature---the modular structure---strongly affects the speed of information diffusion in complex contagion. Our simulations show that there always exists an optimal network modularity for the most efficient spreading process. Beyond this critical value, either a stronger or a weaker modular structure actually hinders the diffusion speed. These results are confirmed by an analytical approximation. 
We further demonstrate that the optimal modularity varies with both the seed size and the target cascade size, and is ultimately dependent on the network under investigation.
We underscore the importance of our findings in applications from marketing to epidemiology, from neuroscience to engineering, where the understanding of the structural design of complex systems focuses on the efficiency of information propagation.

\end{abstract}

\pacs{Valid PACS appear here}
\maketitle


\section{Introduction}

The spread of information in complex networks controls or modulates fundamental processes that can have local effects on individual actors and groups thereof, 
and macroscopic effects on the whole system (e.g., global information cascades). 
Information diffusion has been studied by drawing analogies with epidemics.
Many social behaviors, for example, act like infectious diseases: once triggered, they can spread to the entire population in a very short amount of time, generating a contagion process similar to an epidemic outbreak. 
Examples include collective actions such as voting and participation in social movements, the adoption of innovations such as vaccination and emerging technologies, the diffusion of viral memes in social media, and the spread of norms and cultural fads.
The dynamics of these intriguing and complex phenomena
have attracted research interest from a number of disciplines~\cite{granovetter1973strength, bakshy2012role, cheng2014can}.

There are two major models for the study of information diffusion: the \textit{independent cascade model} and the \textit{linear threshold model}. 
The former assumes that, similar to disease transmission, each exposure is independent from each other and a person only has one chance to ``infect'' their neighbors~\cite{goldenberg2001talk, kempe2003maximizing}. 
The latter postulates that social reinforcement, or exposure to multiple sources, is needed in the contagion process and each person has a threshold to be met for successful adoption~\cite{granovetter1978threshold, watts2002simple}. The independent cascade model suits well with the \textit{simple contagion} scenario, where the goal is to inform people rather than to convince them to take actions~\cite{granovetter1973strength, goldenberg2001talk}. 
It thus has been adopted in the study of word-of-mouth spreading and viral marketing~\cite{, goldenberg2001talk, leskovec2007dynamics}. 
However, some studies revealed that the threshold model is more applicable to the spread of risky or contentious social behaviors for which each additional exposure increases the likelihood of adoption~\cite{watts2002simple, backstrom2006group, centola2007complex, romero2011differences, monsted2017evidence}.
We thus examine the efficiency of information diffusion in the latter case, which is sometimes referred to as \textit{complex contagion}.

Social behaviors spread through social contacts, thus the structure of the underlying social network plays an important role in the process of information diffusion \cite{watts2002simple, onnela2007structure, bakshy2012role, smolla2019cultural}. 
Recent studies have examined the effects of different network properties on the dynamics of information diffusion~\cite{goldenberg2001talk, watts2002simple, galstyan2007cascading}. 

One prominent network feature is \textit{modular structure}---the separation of a network into several subsets of nodes within which connections are dense, but between which connections are sparser~\cite{newman2010networks, fortunato2010community}. 
Networks with many ``bridges'' connecting nodes in different communities tend to have low modularity~\cite{girvan2002community, newman2006modularity}.
Note that we distinguish modularity from another related concept, \textit{clustering}, which refers to the network transitivity and is quantified by the clustering coefficient~\cite{watts1998collective, girvan2002community}. 


The \textit{strength of weak ties} theory suggests that, networks with weak modular structure will promote both the scale and the speed of diffusion since enough shortcuts, that tend to be weak ties, link relatively separated groups and diffuse information across communities~\cite{granovetter1973strength, watts1998collective}.
In contrast, the \textit{weakness of long ties} theory predicts that, in the case of complex contagion where the adoption requires multiple exposures, networks with strong modular structure, and thus an abundance of strong ties, can enhance the spread of certain social behaviors~\cite{centola2007complex, centola2010spread}. 
The two competing hypotheses based on prior theoretical work manifest the interplay between social reinforcement and network modularity in most real social networks.
Yet, empirical studies seem to reveal inconsistent results regarding the role played by community structure in complex contagion~\cite{centola2010spread, weng2013virality, romero:icwsm2013}.
Recent findings reveal that network modularity plays two different roles in information diffusion, namely (i) enhancing intra-community spreading, and (ii) hindering inter-community spreading~\cite{nematzadeh2014optimal}, providing an in-principle unifying explanation to the competing empirical evidence.

Overall, prior work on the relationship between network modularity and large cascades has mainly focused on one aspect of information diffusion---the size of information cascades, \textit{i.e.}~the total number of ``infected'' individuals in the steady state. Another important cascade feature---the efficiency of information diffusion, \textit{i.e.}~the total time it takes to reach the steady state---has been underexplored~\cite{iribarren2009impact, karsai2011small, delvenne2015diffusion}.
A better understanding of information diffusion speed can have many practical applications, such as informing the design of communication networks where the efficiency of information flow needs to be prioritized. 
For instance, the adoption of preventive measures such as wearing masks and social distancing during the COVID-19 pandemic may need to be optimized for adoption speed in order to ``flatten the curve''.

The extant literature has also demonstrated how insights about the interplay between network modularity and information spread can provide a principled understanding of various complex system dynamics, from characterizing neuronal communication in human connectomics~\cite{mivsic2015cooperative}, to optimizing immunization strategies for public health and animal welfare~\cite{yan2015global, sah2017unraveling}.

\section{Models}

\subsection{Diffusion Model}

Here, we systematically examine the effects of network modularity on the \emph{speed} of information diffusion in complex contagion by utilizing the linear threshold model~\cite{granovetter1978threshold, watts2002simple}.
We define diffusion speed as the average rate of a spreading process, measured as the eventual growth of the cascade divided by the time it takes to reach equilibrium.
We show that, in complex networks, there always exists an optimal amount of modularity for the most efficient information diffusion process.

In the linear threshold model, a node can be in two states: either active or inactive. 
Each node $a$ is assigned a threshold $\theta_a$ uniformly at random from the interval $[0, 1]$. 
Initially all nodes are inactive.
At time step $t=0$, a fraction $\rho_0$ of $N$ nodes (the seeds) are switched into active state.
In the subsequent time steps, a node can become active if its fraction of active neighbors exceeds the threshold, and it stays active forever once being activated.
Following these rules, we update a fraction $f$ of all nodes (selected randomly) at each step.
In the synchronous updating scenario, where $f=1$, the contagion process unfolds in a deterministic manner until the network reaches the steady state \cite{watts2002simple, kempe2003maximizing, nematzadeh2014optimal}.
This model can be adapted to the case of asynchronous updating by setting $f<1$.
We assume that all nodes have the same threshold $\theta$ \cite{singh2013threshold, nematzadeh2014optimal}.
We measure the time steps $t_s$ it takes to reach steady state and the total fraction $\rho_{t_s}$ of active nodes across the network at $t_s$. The average speed of diffusion is: $\bar{v}=(\rho_{t_s}-\rho_0)/t_s$.

\subsection{Network Model}

We adopt the stochastic block model (SBM) to generate networks with community structure~\cite{karrer2011stochastic}.
The underlying network consists of $N$ nodes partitioned into $d$ communities $\{C_1,C_2,...,C_d\}$. 
Let $|C_i|$ be the size of $C_i$, and $\rho^{(i)}_{t}$ be the fraction of active nodes in $C_i$ at time $t$.
Each community $C_i$ has a specified degree distribution $p_{k}^{(i)}$ and a mean degree $z^{(i)}=\sum kp_{k}^{(i)}$.
The edges in the network are randomly distributed according to a $d \times d$ mixing matrix $\mathbf{e}$, with $e_{ij}$ defined as the fraction of edges that connect nodes in $C_i$ to nodes in $C_j$.
Although studies have indicated that tie strength is an important factor in modeling information diffusion~\cite{goldenberg2001talk, onnela2007structure}, here we consider edges to be unweighted, due to the unclear relationship between tie strength and network topology---some studies argue that strong ties mostly reside within tightly knit clusters and weak ties tend to link together distant communities \cite{granovetter1973strength, goldenberg2001talk, centola2007complex, onnela2007structure}, while other empirical work reveals the opposite conclusion in social and scientific collaboration networks \cite{de2014facebook, ke2014tie, petersen2015quantifying}.

\subsection{Numerical Simulation}

We use numerical simulations to compare the speed of diffusion across an ensemble of networks with different strength of network modularity.
For simplicity, here we consider the case of two equally sized communities: let $d=2, |C_1| = |C_2| = N/2$, and the seed nodes are randomly selected from $C_1$, thus $\rho_0^{(1)} = 2\rho_0, \rho_0^{(2)} = 0$.
We assume $p_{k}^{(1)}$ and $p_{k}^{(2)}$ both follow a Poisson distribution, with $z^{(1)} = z^{(2)} = z$. 
The expected total number of edges is: $M = zN/2$.
Let $\mu M$ edges be randomly distributed between $C_1$ and $C_2$, and the remaining $(1-\mu) M$ edges be randomly placed between node pairs in the same community, thus $\mathbf{e} = \frac{1}{2}\lbrack\begin{matrix}1-\mu, & \mu\\ \mu, & 1-\mu\end{matrix}\rbrack$.
Here $\mu$ controls the strength of network modularity which turns out to be $Q = 1/2 - \mu$, based on the current partition.
A larger $\mu$ gives a network with weaker network modularity since there are more edges running between two communities.
For each $\mu$, we run $100$ simulations, with each assuming a different realization of the network and the seeds.


\subsection{Analytical Approximation}

We also study the dynamics in our system analytically. 
The cascade size $\rho_t$ is equal to the probability that a randomly chosen node is active at time $t$.
The topology of such a large network can be approximated by a tree structure with infinite depth and a single node at the top, a.k.a. a \textit{tree-like approximation} ~\cite{gleeson2008cascades}.
The top node is connected to $k_a$ neighbors at the next lower level, while any other node $a$ at level $n$ is connected to $k_a-1$ neighbors at level $n-1$, where $k_a$ is the degree of node $a$. 
At any level, the probability that a node in $C_i$ is among the seeds is $\rho_0^{(i)}$.
In synchronous updating, the tree level $n$ can be directly mapped to the time step $t$ used in simulations~\cite{gleeson2008cascades}, which means that $\rho_{t}$ can be approximated as the probability $\rho_n$ that the top node is active, assuming that it resides at level $n=t$, since the top node can only be infected by nodes at most $n$ levels below.
We can calculate the probability of being active from nodes at the bottom level ($n=0$) to the top node ($n=t$), one level at a time, according to the linear threshold model.
See the derivation of $\rho_n$ in \textit{Materials and Methods}..

\begin{figure}[t]
\centering
\includegraphics[trim=0mm 0mm 0mm 0mm, width=\columnwidth]{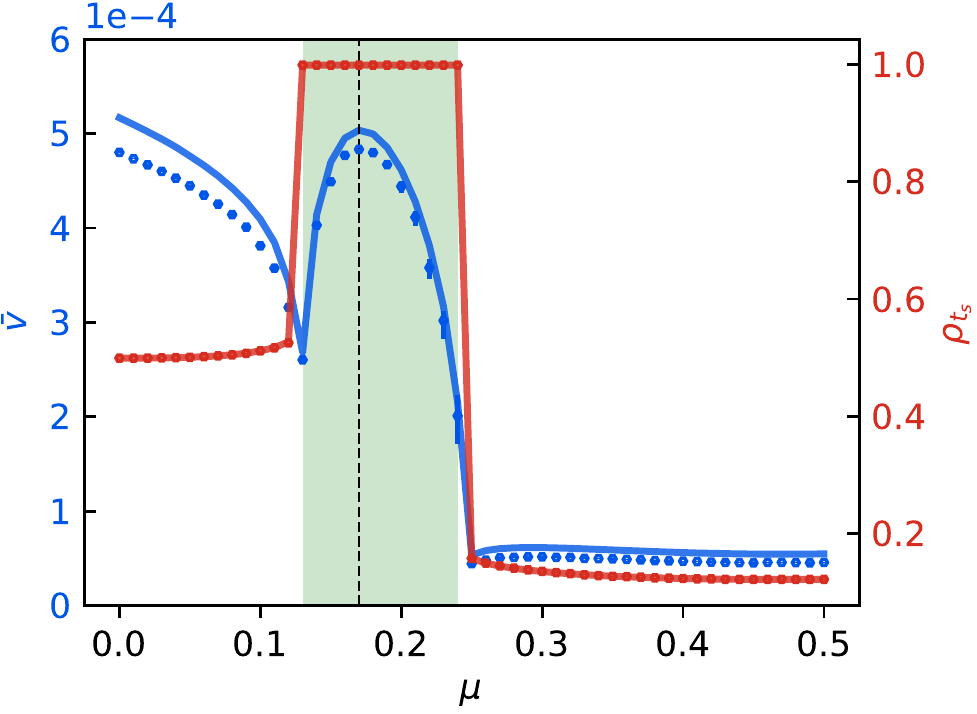}
\caption{Simulation results (dots) and analytical predictions (lines) of Eqs.~\ref{avg-speed} (cf., \textit{Materials and Methods}). 
\textbf{Blue axis}: the average speed of information diffusion, $\bar{v}$. \textbf{Red axis}: the size of information cascade, $\rho_{t_s}$. 
\textbf{Green area}: the range of $\mu$ that can enable global cascades ($\rho_{t_s} \geqslant 0.99$).
The $x$-axis represents the strength of network modularity controlled by $\mu$. 
The dashed vertical line corresponds to $\mu = 0.17$ that yields the highest $\bar{v}$ by prediction.
The simulation results are averaged over $100$ realizations of the network for each $\mu$, with $N=1\times10^5, z=10, \rho_0=0.1, \theta=0.35, f=0.01$.
The error bars indicate the interquartile ranges.}
\label{fig-one}
\end{figure}

\section{Results}

\subsection{Optimal Modularity for the Speed of Global Cascades}

Fig.~\ref{fig-one} displays an interval of network modularity that can trigger global cascades, which concurs with the findings in~\cite{nematzadeh2014optimal}. 
Intuitively, one would imagine that a stronger modularity (smaller $\mu$) increases diffusion speed in $C_1$ since nodes in $C_1$ are exposed to more seeds, while a weaker modularity (larger $\mu$) increases diffusion speed in $C_2$ because more bridges connect nodes in $C_2$ to the seeds.
This observation, however, raises the following question: is there an ideal network modularity at which the global cascade reaches the highest average diffusion speed?


Let us first analyze the behavior of our system when only local diffusion is possible. Fig.~\ref{fig-one} indicates that, when the network modularity is too strong (very small $\mu$), information only spreads among nodes in $C_1$ due to the lack of bridges between two communities, thus decreasing modularity (increasing $\mu$) decreases the average diffusion speed because it takes longer for spreading in $C_1$ and the cascade size stays the same.

When a global cascade is achieved, however, there is a quadratic relationship between  the average diffusion speed and network modularity: decreasing modularity first increases the average diffusion speed, but only up to a critical point, after which a further reduction in modularity slows down the overall diffusion dynamics. 
The global cascade thus reaches its highest average speed at the optimal network modularity ($\mu = 0.17$). 
The analytical predictions show excellent agreement with the simulations (Fig.~\ref{fig-one}).

\begin{figure}[t]
\centering
\includegraphics[trim=0mm 0mm 0mm 0mm, width=\columnwidth]{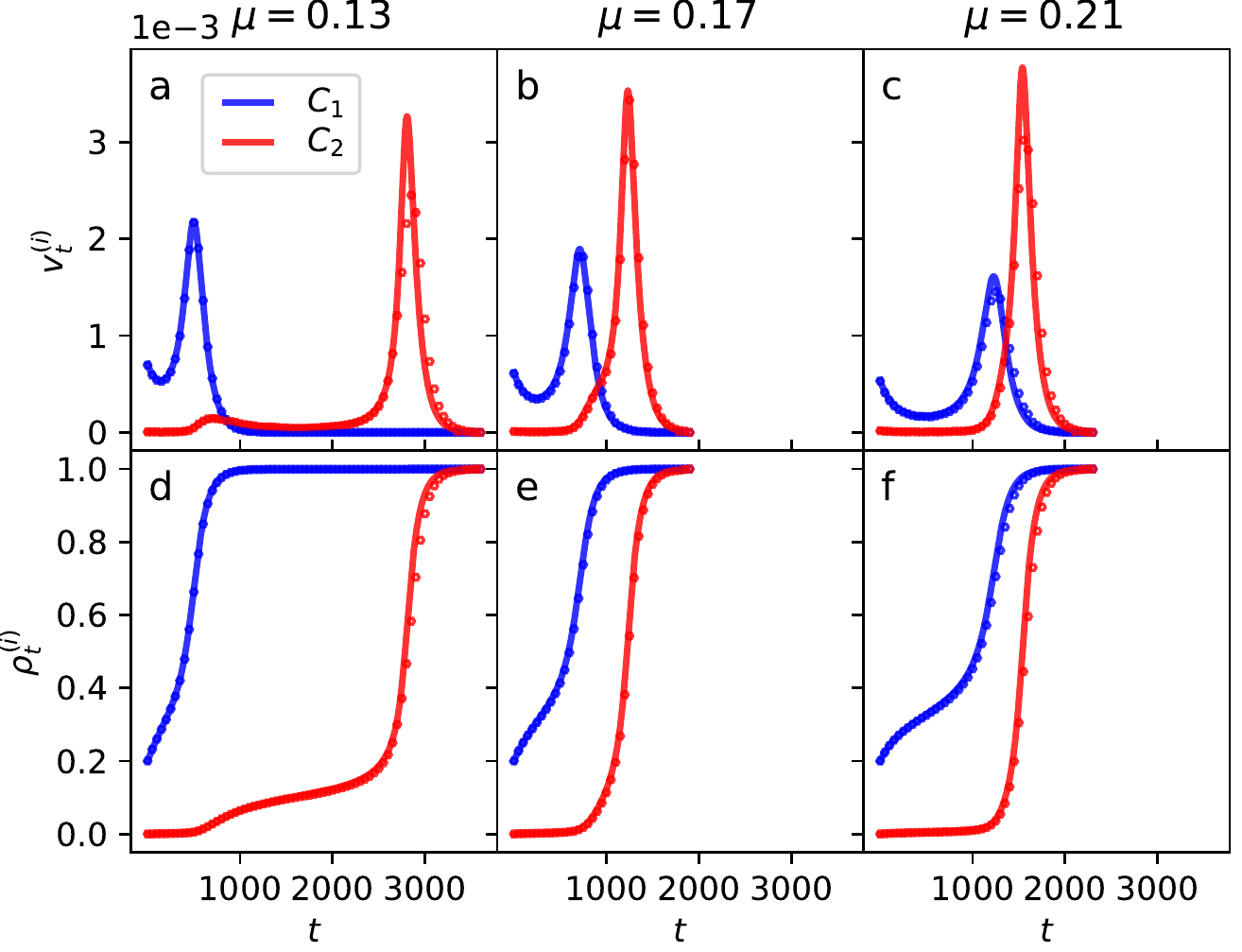}
\caption{Cross sections of three different $\mu$ values in Figure~\ref{fig-one} that enable global cascades. 
(a-c) The diffusion speed $v^{(i)}_t$ in $C_1$ and $C_2$ as a function of time step $t$. 
(d-f) Same as (a-c), but for the cumulative cascade size $\rho^{(i)}_t$. 
The theoretical predictions of Eqs.~\ref{step-speed} (lines) show excellent agreement with the numerical simulations (dots), averaged over $100$ runs. 
The optimal $\mu = 0.17$ achieves the shortest total diffusion time, thus the highest average diffusion speed.}
\label{fig-two}
\end{figure}


Next, we analyze the cascade dynamics in more detail to understand this phenomenon. 
Fig.~\ref{fig-two} shows the diffusion speed per time step in each community, for three different levels of network modularity.
The time lags of spreading in two communities can help us to explain the influence of network modularity on the average diffusion speed of global cascades.

At $\mu = 0.13$, we reach the lower bound of the window for global cascades. 
However, the time difference between $C_1$ and $C_2$ is the longest: the spreading in $C_2$ merely gets started after $C_1$ reaches steady state (Fig.~\ref{fig-two}a). 
Thus the relatively long diffusion time in $C_2$ is the bottleneck for the average diffusion speed at global scale.

One may, therefore, predict that the highest average diffusion speed can be achieved when the time lag between the two communities is reduced as much as possible.
For instance, since the time difference to finish spreading at $\mu=0.21$ (Fig.~\ref{fig-two}c) is shorter than that at $\mu=0.17$ (Fig.~\ref{fig-two}b), the average diffusion speed would be predicted to be faster in the former case (Fig.~\ref{fig-two}c).
However, such an inference is incorrect, as the diffusion at $\mu=0.21$ takes longer time than the scenario when $\mu=0.17$, for which the global cascade finishes in the shortest amount of time.

Comparing the optimal network modularity (Fig.~\ref{fig-two}b) to the first scenario (Fig.~\ref{fig-two}a), it takes slightly more time to finish spreading in $C_1$, due to the decreasing number of edges in $C_1$.
But the increasing connections between the two communities reduces the diffusion time in $C_2$.
The time lag between $C_1$ and $C_2$ is much shorter, but not close to zero.
Fig.~\ref{fig-two} indicates that, at this optimal network modularity, neither $C_1$ nor $C_2$ achieves its highest diffusion speed, but both are pretty close to it, resulting in the most efficient global cascade.

However, at $\mu=0.21$, the further reduction of the number of edges in $C_1$ slows down the speed of local spreading in $C_1$, and this becomes the bottleneck of the average speed at global scale. 
Although, under this condition, $C_1$ and $C_2$ reach the steady state almost concurrently (with a time lag close to zero), it cannot counteract the increase in diffusion time for both communities (Fig.~\ref{fig-two}c).


\begin{figure}[t]
\centering
\includegraphics[trim=0mm 0mm 0mm 0mm, width=\columnwidth]{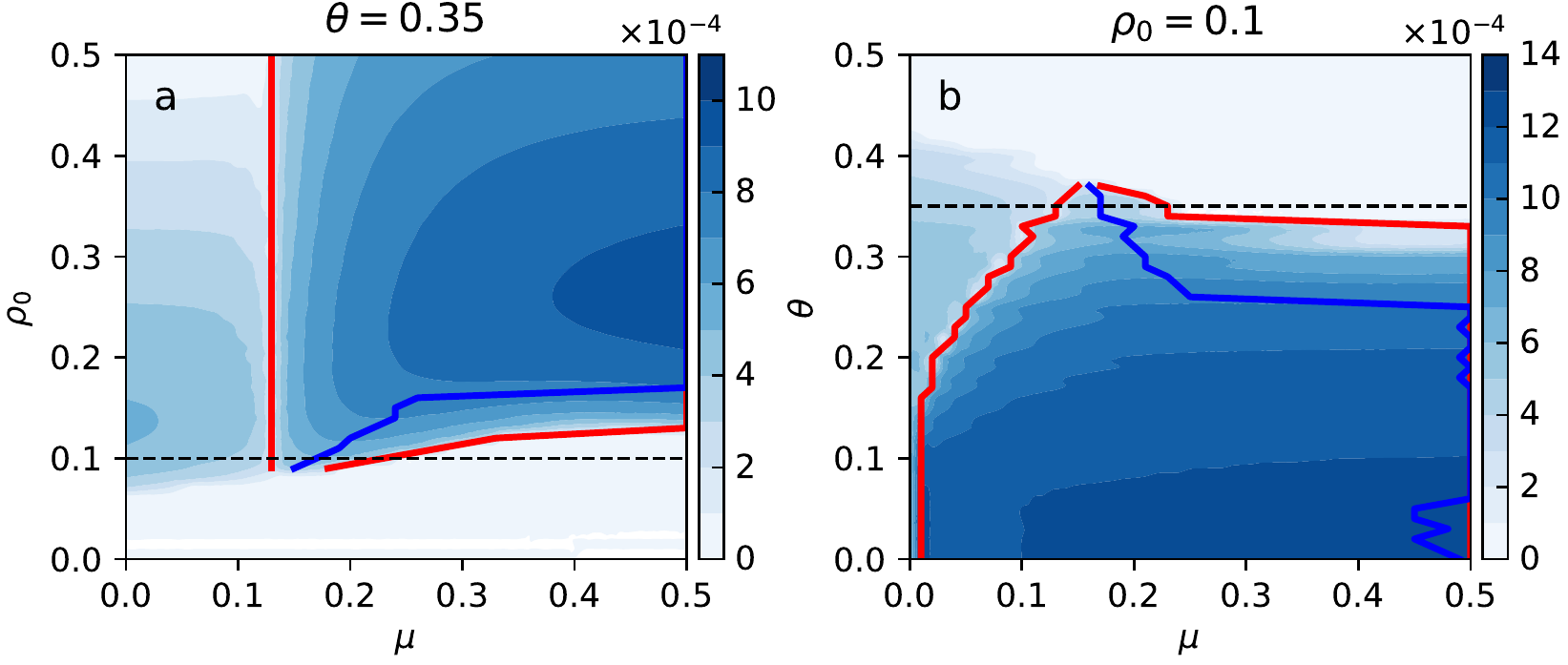}
\caption{Phase diagrams of the average diffusion speed $\bar{v}$ as a function of seed size $\rho_0$ (left) and threshold $\theta$ (right) on SBM networks.
The two red curves mark the region for global cascades. 
The blue curve represents the $\mu$ value that yields the highest $\bar{v}$ for a given $\rho_0$ or $\theta$. 
The results are averaged over $100$ simulations for each combination of $(\rho_0, \mu)$ or $(\theta, \mu)$.
Simulation parameters are: $N=1\times10^5, z=10, f=0.01$, with $\theta=0.35$ (left) and $\rho_0 = 0.1$ (right).
The seeds are randomly selected from a single community.
The dashed line is a slice at $\rho_0=0.1$ ($\theta=0.35$) in Figure~\ref{fig-one}.}
\label{fig:phase}
\end{figure}

\subsection{The Effects of Seed Size and Threshold}

Fig.~\ref{fig:phase} presents two phase diagrams of the average diffusion speed $\bar{v}$ as a function of the seed size $\rho_0$ and the threshold $\theta$. 
It indicates that, in the region of global cascades, there always exists an optimal modularity for the most efficient information diffusion, and this critical value of $\mu$ depends on both $\rho_0$ and $\theta$.

Fig.~\ref{fig:phase}a shows that a minimal seed size is needed to trigger global cascades, and once above this threshold, when $\rho_0$ is not too large (e.g., $\rho_0=0.1$), the average speed of global cascades first increases and then decreases as one reduces the modularity (increasing $\mu$), resulting in an intermediate value of $\mu$ as the optimal modularity. However, when $\rho_0$ is sufficiently large (e.g., $\rho_0=0.2$), the average speed of global cascades always increases as one increases the number of cross-community links, making the network with no-community structure ($\mu=0.5$) the ideal case for the most efficient spreading process. This can be explained by the fact that, when increasing $\mu$ never blocks local spreading in $C_1$ as a result of the presence of enough seeds in $C_1$, more external links are always going to make the diffusion faster in $C_2$. Similar patterns emerge for the threshold $\theta$ when the seed size is fixed (Fig.~\ref{fig:phase}b).

We obtain consistent results on SBM networks with different network sizes, variable average degrees, different seed arrangements between $C_1$ and $C_2$, and arbitrary number of equally sized communities, based on both simulations and analytical predictions (\si{Section \RomanNumeralCaps{3}, A-E}).

\subsection{Simulations on non-SBM Networks}

\begin{figure}
\centering
\includegraphics[trim=0mm 0mm 0mm 0mm, width=\columnwidth]{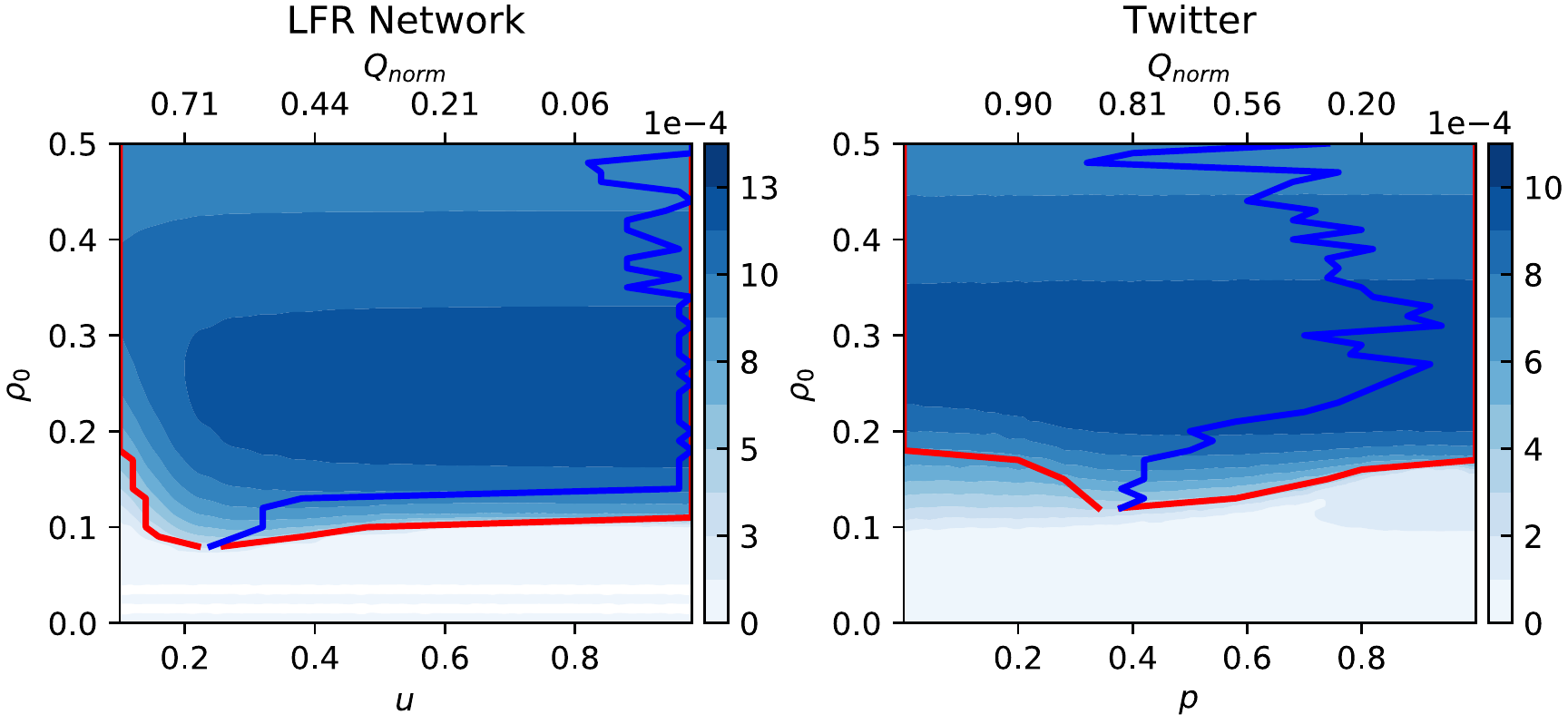}
\caption{Phase diagrams of the average diffusion speed $\bar{v}$ on the LFR (left) and Twitter (right) networks.
Parameters $\mu$ and $p$ on the $x$-axis control the network modularity.
The blue curve indicates the optimal $\mu$ (or $p$) for $\bar{v}$ for a given seed size $\rho_0$.
The normalized modularity $Q_{\text{norm}}$ with respect to $\mu$ (or $p$) is shown on the top axis.
Network statistics are: $N=25000, z=10, \gamma=2.5, \beta=1.5, k_{max}=30$ (for LFR); $N=81306, z=16$ (for Twitter). 
Simulations are averaged over $100$ runs, with $\theta=0.3, f=0.01$. The seeds are randomly selected across the whole network.}
\label{fig:lfr-twitter}
\end{figure}

Although the SBM provides reproducible and well controlled modular networks for modeling the speed of information diffusion using tractable computational approaches, it is clearly appealing to test the generalizability of our findings on networks without the assumptions of equally sized communities and randomly distributed edges. 
To this end, we perform simulations on networks with more complex structure, such as heterogeneous communities, high clustering, and power-law degree distribution.
We also randomly select the seed nodes across the network, instead of placing them in a single community.

We use the LFR benchmark graph to generate synthetic networks with community structure similar to that observed in real-world networks~\cite{lancichinetti2008benchmark}.
We also simulate information diffusion on a Twitter network (\textit{Materials and Methods}) and six other real-world networks (\si{Section \RomanNumeralCaps{3}, F}).

The phase diagrams for both types of networks are shown in Fig.~\ref{fig:lfr-twitter}.
An optimal modularity for the most efficient global cascades still emerges as in the case of SBM networks (Fig.~\ref{fig:phase}a).
Fig.~\ref{fig:lfr-twitter} shows that the minimal seed size required to trigger global cascades depends on the network under investigation: a $10\%$ random sample of all nodes is enough to generate global diffusion on LFR networks for a wide range of modularity, while the same fraction of seed nodes only generates small cascades on the Twitter network.
For the same reason, the optimal normalized modularity (\textit{Materials and Methods}) for a given seed size also changes across networks.
Different from the SBM, when the seed size is large enough, small changes in modularity only result in small changes in the average diffusion speed since the diffusion tends to reach global cascades rapidly due to the fact that seeds are randomly distributed over the whole network. Thus, for large seed sizes, the modularity for the fastest global cascades fluctuates on both LFR and Twitter networks, as opposed to the case of SBM networks, where the optimal values are always the same (Fig.~\ref{fig:phase}a).
This also explains why the position of the phase transition at which global cascades emerge moves to the highest modularity for large seed sizes (Fig.~\ref{fig:lfr-twitter}).

Overall, the optimal modularity is ultimately dependent on the network under investigation. This is because their overall network structure is very different from each other, such as the degree distribution, the clustering coefficient, the community sizes, etc., and the interactions between modularity and these network properties can greatly impact diffusion dynamics. However, the general trend---that the optimal modularity decreases as the seed size increases---is preserved for a variety of complex networks (\si{Section \RomanNumeralCaps{3}, F}).

\subsection{Optimal Modularity for Different Cascade Sizes}

\begin{figure*}[t]
\centering
\includegraphics[trim=0mm 0mm 0mm 0mm, width=\linewidth]{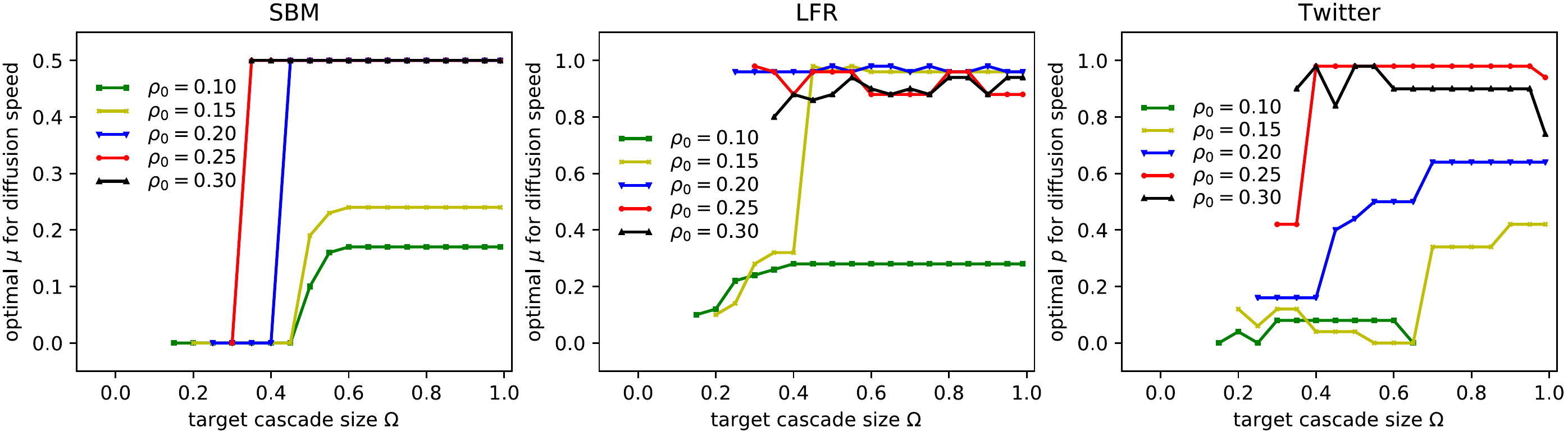}
\caption{The optimal network modularity for fast information diffusion changes as a function of the target cascade size on three different types of networks.
The modularity is controlled by $\mu$ for SBM and LFR networks (by $p$ for the Twitter network).
The optimal values of $\mu$ and $p$ are selected from those that can achieve the given target size $\Omega$.
All simulation parameters are the same as in Fig.~\ref{fig:phase}a and Fig.~\ref{fig:lfr-twitter}.
Note that a small $\rho_0$ may not be able to reach all target cascade sizes, e.g., a seed size of $\rho_0=0.1$ on the Twitter network is only able to infect up to about $\Omega = 65\%$ of all nodes. 
Line plots for different $\rho_0$ start from different $\Omega$ since $\Omega > \rho_0$.}
\label{fig:mu-vs-target-size}
\end{figure*}

The objective of certain diffusion scenarios is not always to reach the global cascades. 
For instance, in the case where an organization needs to get at least $x$ signatures among its members before a certain date in order to get an initiative on a ballot, the goal is to activate just a fraction of the whole population.
This example prompts us to ask: how does the optimal modularity for speed change with different target cascade sizes?



To answer this question, for a fixed $\Omega$ (i.e., the target cascade size) and $\rho_0$ (i.e., the seed size), we determine the optimal value of $\mu$ (or $p$) that minimizes the time it takes for the cascade to reach $\Omega$.
Fig.~\ref{fig:mu-vs-target-size} indicates that the optimal modularity for the average diffusion speed typically decreases as the target cascade size $\Omega$ increases. For instance, the optimal $\mu$ changes from $\mu=0$ to $\mu=0.17$ as $\Omega$ increases from $\Omega=0.15$ to $\Omega=0.99$ for $\rho_0=0.1$ on SBM networks. 
The intuition behind this result is that since the originating communities already contain enough nodes to satisfy the small target cascade size, it is better to have strong modularity to facilitate local spreading (Fig.~\ref{fig-one}).
However, when the seed size is large enough (e.g., $\rho_0=0.3$), the optimal modularity tends to be small for both large and small (relative to $\rho_0$) target cascades. In this case, originating communities can quickly be saturated and thus the best strategy is to promote large cascades through inter-community edges.


This observation provides a more complete picture of our findings: the best network to optimize diffusion speed is not always the same, suggesting that the target cascade size, together with the seed size, should all be taken into consideration when designing the most efficient network.

Beyond a constraint on the cascade size, there are other situations where one needs to optimize the diffusion speed with a time budget (or equivalently to maximize the cascade size in a given time window). We thus further examine the diffusion dynamics by considering having a limit on the diffusion time in \si{Section \RomanNumeralCaps{3}, G}, which shows that the optimal modularity tends to decrease as the time budget increases.

\section{Discussion}

We investigate the effect of community structure, as measured by network modularity, on the speed of information diffusion. 
Through simulations and analytical approximations, we reveal that there always exists an optimal strength of modularity---under which information or behavior diffuses at the highest average speed. 
We demonstrate that such an efficient spreading behavior is achieved by making the right compromise between internal connectivity and cross-community bridges for synchronized diffusion in different communities.
We also find that the optimal modularity varies with respect to the seed size and the target cascade size. These findings are consistent on both synthetic and real-world networks.

Our findings provide insights for many real-world applications that allow for the optimization of network structure to enable rapid diffusion or adoption. 
For instance, it may help to design better organizational structure for firms with many different functional departments where the efficiency of diffusion is important (e.g., the adoption of social norms and work habits such as working hard). Drawing on the communication network of employees in a company (e.g., from email or social media), managers could make office assignments as an intervention to help change the interaction patterns such that the network approaches its optimal modularity, and thus making the process of social contagion more efficient.

In relation to epidemics, our findings inform how to speed up the adoption of preventive health measures in order to slow down the spread of infectious diseases. In the case of the COVID-19 pandemic, for example, one can optimize the modularity of social networks at different levels to enable the rapid adoption of healthy behaviors such as wearing face masks, staying at home, and practicing social distancing, in order to reduce virus transmission and disease spread.
In preventative health, one intervention used by practitioners to address lifestyle-related public health challenges like obesity is to modify the contact network of a community to promote the spread of healthy behaviors, such as by providing role models or ``health buddies'' to mothers, young children, and users in online health communities~\cite{centola2010spread, salvy2017home, wilder2018optimizing}. Our finding suggests that the modularity should be taken into consideration in the network modification procedure to maximize the speed of behavioral change.

Online networks can be reshaped to influence information diffusion dynamics: social media platforms, for example, can design their friend recommendation algorithms to change the network modularity to promote (e.g., advertisements) or suppress (e.g., participation in illicit activities) diffusion processes.


Although our study is postulated upon the premise that one can alter network structures to maximize diffusion speed, our findings still have implications for real-world networks with a structure that cannot be modified: one can quantify the degree of efficiency the network is functioning at and determine the optimal seed size for a given network and diffusion process.


This study also has implications for online campaigns. Social media users often receive content relevant to their interests in trending discussions or ephemeral events. Our study suggests that advertisers can target networks with a high level of modular structure to maximize the campaign reach and inform a large audience in a short period of time. For instance, a petition to the White House that needs to gather $100,000$ signatures in just $30$ days can be promoted within high-modularity social networks to increase the chance of success.

From a methodological standpoint, by incorporating the effect of network modularity on the diffusion speed, machine learning algorithms can utilize modularity to better predict the efficiency of information cascades.
Our framework can allow the study of many naturally-occurring complex systems in biological networks, and enable the understanding of evolutionary dynamics in complex networks exhibiting a certain level of modularity that facilitates or hinders diffusion speed. For example, network modularity has already been used to study spreading dynamics on the human connectome and to explain global communication on brain networks~\cite{mivsic2015cooperative}, but the communication speed in this context is unexplored.

Future work can focus on the empirical validation of the relationship between network modularity and the efficiency of information diffusion, and on examining its variations by considering other diffusion mechanisms (e.g., the independent cascade model) on networks with even more complex structure such as the hierarchical organization of communities.

\section{Materials and Methods}

We provide the updating equations of the analytical approximation, the details of the network data, and the normalized network modularity measure below.

\subsection{The Calculation of Diffusion Speed}

Note that the tree-like approximation only deals with probabilities; it does not represent the actual diffusion process on a particular network, where the spreading always starts from the seeds, not from nodes at the bottom level. Here, $\rho_{n} = \sum\nolimits_{i}\rho_{n}^{(i)}|C_{i}|/N$, with $i$ in $\rho_{n}^{(i)}$ indicating that the top node at level $n=t$ belongs to $C_i$ and $|C_{i}|$ is the size of each community.
We can iteratively calculate the cascade size using the following updating equations (see \si{Section \RomanNumeralCaps{2}} for details):

\begin{align}
    \bar{q}_{n}^{(i)} &= \frac{\sum\nolimits_{j}e_{ij}q_{n-1}^{(j)}}{\sum\nolimits_{j}e_{ij}} = \frac{1}{d}\sum\nolimits_{j}e_{ij}q_{n-1}^{(j)},\\
    q_{n}^{(i)} &= \rho_0^{(i)} + (1-\rho_0^{(i)}) \sum_{k}\tilde{p}_{k}^{(i)}\sum_{m=\ceil*{\theta k}}^{k-1}\binom{k-1}{m}(\bar{q}_{n}^{(i)})^m (1-\bar{q}_{n}^{(i)})^{k-1-m},\\
    \rho_{n}^{(i)} &= \rho_0^{(i)} + (1-\rho_0^{(i)}) \sum_{k}p_{k}^{(i)}\sum_{m=\ceil*{\theta k}}^{k}\binom{k}{m}(\bar{q}_{n}^{(i)})^m (1-\bar{q}_{n}^{(i)})^{k-m}.
\end{align}
The diffusion speed in $C_i$ at time $t$ can be approximated as 
\begin{align} \label{step-speed}
    v_t^{(i)}=d\rho_{t}^{(i)}/dt = [\rho_{t+1}^{(i)}-\rho_{t}^{(i)}]^{+},
\end{align}
where the notation $[\cdot]^{+}$ stands for $\max(0, \cdot)$. 
The overall diffusion speed $v_t$ at time $t$, the total diffusion time $t_s$, and the average diffusion speed $\bar{v}$ are

\begin{align} \label{avg-speed}
    v_t = \sum_{i}\frac{|C_{i}|}{N}v_t^{(i)}, \qquad
    t_s = t \mid v_t = 0, \qquad
    \bar{v} = \frac{\rho_{t_{s}} - \rho_{0}}{t_s}.
\end{align}

\subsection{LFR Network}
The node degrees and community sizes in LFR networks both follow a power law distribution, with exponents $\gamma$ and $\beta$, respectively.
The typical values of the exponents are: $2\leqslant\gamma\leqslant3$, $1\leqslant\beta\leqslant2$. Here, we let $\gamma=2.5, \beta=1.5$.
Similar to SBM, LFR networks also use a parameter $\mu$ to control for the modularity, which is defined as the fraction of a node's edges to others outside its community.
Unlike SBM, the node partition in LFR networks is not fixed for different $\mu$, and $0\leqslant\mu\leqslant1$ since the number of communities is typically larger than $2$ (\si{Section \RomanNumeralCaps{3}, C\&E}).

\subsection{Twitter Network}
The Twitter network data is obtained from~\cite{snapnets}. 
The largest connected component (LCC) of its undirected network consists of 81K nodes and 1.3M edges. We detect 70 communities for the LCC using the Louvain algorithm~\cite{blondel2008fast}.

We rewire edges to change the network modularity. 
For each rewire, we do the following: (i) with probability $p$, we randomly select a pair of communities and randomly select a within-community edge from each community, and then swap the edge ends if it is possible (no parallel edges are allowed); (ii) with probability $(1-p)$, we randomly select a pair of communities and randomly select two cross-community edges running between them, and then swap the edge ends to create two within-community edges if it is possible. 
The above process is repeated about 650K times so that each edge can be rewired once on average.
Parameter $p$ here is similar to $\mu$ used in SBM and LFR networks: a small $p$ increases the modularity, while a large $p$ decreases the modularity.

\subsection{Normalized network modularity}
The network modularity $Q$ quantifies the number of intra-community edges minus the expected number if edges are placed at random, for a given node partition.
It achieves the maximum value $Q_{\text{max}}$ on a perfectly mixed network where all edges connect nodes in the same community. However, $Q_{\text{max}}$ typically varies from network to network.
To compare the strength of modularity across different networks, we therefore use the normalized value of the modularity: $Q_{\text{norm}} = Q/Q_{\text{max}}$~\cite{newman2010networks}. Note that, in Figs.~\ref{fig-one}-\ref{fig:phase}, $Q = 1/2 - \mu$, $Q_{\text{max}}=1/2$, and $Q_{\text{norm}}=1-2\mu$. For LFR (Twitter) networks, the relationship between $Q_{\text{norm}}$ and $\mu$ ($p$) is nonlinear, as shown in Fig.~\ref{fig:lfr-twitter}.\\

\subsection{Materials and data availability} 
Data for all seven real-world networks used in this study is available at: \url{https://snap.stanford.edu/data/}. 
A public repository with code to reproduce our results is available at: \url{https://github.com/haoopeng/diffusion_speed}.
\\

{\footnotesize
\textbf{Author contributions}
We thank Aparna Ananthasubramaniam, Ceren Budak, Ashok Deb, Danaja Maldeniya, and Ed Platt for helpful discussions and suggestions. This work is partly supported by DARPA (W911NF-17-C-0094) and by the Air Force Office of Scientific Research under award number FA9550-19-1-0029.
}

{\footnotesize
\textbf{Author contributions}
H.P., A.N., D.M.R. and E.F. collaboratively conceived and designed the study. H.P. carried out the experiments and performed the analyses. H.P., D.M.R. and E.F. drafted and revised the final manuscript.
}

\bibliographystyle{plain}
\bibliography{main}
\balance

\end{document}


\title{Network Modularity Controls the Speed of Information Diffusion\\(Supplementary Information)}

\author{Hao Peng}
\affiliation{School of Information, University of Michigan}
\author{Azadeh Nematzadeh}
\affiliation{S\&P Global}
\author{Daniel M. Romero}
\affiliation{School of Information, University of Michigan}
\author{Emilio Ferrara}
\affiliation{Information Sciences Institute, University of Southern California}
\date{\today}

\maketitle

\section{Introduction}

The network representation of social relationships between people is a core ingredient in modeling the dynamics of information diffusion since the adoption of ideas or social behaviors are often influenced by one's social neighbors. 
Therefore the structure of the underlying social network strongly affects the process of information diffusion. 
In this paper, we study how a salient network property---the modular structure---influences the speed of information diffusion by using the linear threshold diffusion model on networks with varying degree of network modularity.
Through both simulations and an analytical approximation, we demonstrate that there exists an optimal network modularity for the most efficient information diffusion at global scale.

In this supplementary document, we present further evidence to support our findings by examining the behavior of our diffusion model under more general conditions with a wide range of parameters.
We investigate the average speed of information diffusion on SBM networks with varying (i) network size $N$, (ii) average degree $z$, (iii) anti-modular structure, (iv) seed arrangements, and (v) number of communities $d$. 
Additionally, we report results based on many real-world networks where the seed nodes are randomly selected across the whole network instead of from a single community.
Finally, we present qualitatively similar results by considering a different constraint---fixing the diffusion time instead of fixing the cascade size---in measuring the average diffusion speed.

\section{The Tree-like Approximation of Diffusion Speed}

As discussed in the main paper, $\rho_{n} = \sum\nolimits_{i}\rho_{n}^{(i)}|C_{i}|/N$, with $i$ in $\rho_{n}^{(i)}$ indicating that the top node belongs to $C_i$ and $|C_{i}|$ is the size of each community.
To calculate $\rho_{n}^{(i)}$, we introduce two auxiliary variables: $q_n^{(i)}$ and $\bar{q}_{n}^{(i)}$. Let $q_n^{(i)}$ be the probability that a node in $C_i$ at level $n$ is active, conditioning on its parent being inactive, and $\bar{q}_{n}^{(i)}$ be the probability of reaching an active child at level $n-1$ by following an edge from an inactive node in $C_i$ at level $n$. We can update $q_n^{(i)}$ and $\bar{q}_{n}^{(i)}$ using Eq.~\ref{q-bar}-\ref{q-func}:

\begin{align}
    \bar{q}_{n}^{(i)} = \frac{\sum\nolimits_{j}e_{ij}q_{n-1}^{(j)}}{\sum\nolimits_{j}e_{ij}} = \frac{1}{d}\sum\nolimits_{j}e_{ij}q_{n-1}^{(j)},
\label{q-bar}
\end{align}

\begin{equation}
\begin{gathered}
    q_{n}^{(i)} = \rho_0^{(i)} + (1-\rho_0^{(i)}) \sum_{k}\tilde{p}_{k}^{(i)}\sum_{m=\ceil*{\theta k}}^{k-1}\binom{k-1}{m}(\bar{q}_{n}^{(i)})^m \\
    \times (1-\bar{q}_{n}^{(i)})^{k-1-m} \equiv g^{(i)}(\bar{q}_{n}^{(i)}),
\end{gathered}
\label{q-func}
\end{equation}
where $\tilde{p}_{k}^{(i)}$ is the probability that a node in $C_i$ reached by following an edge from its inactive parent has degree $k$, thus $\tilde{p}_{k}^{(i)} = kp_{k}^{(i)}/z^{(i)}$~\cite{newman2010networks}. Note that $q_{0}^{(i)} = \rho_{0}^{(i)}$.
Eq.~\ref{q-func} is the sum of two scenarios: (i) the probability that the node is among the seeds ($\rho_0^{(i)}$), and (ii) the probability that the node is not among the seeds ($1 - \rho_0^{(i)}$) but is connected to at least $\ceil*{\theta k}$ active children (the second summation, note that this node connects to $k-1$ children), summed over all possible degrees $k$ of that node (the first summation).

Similar to $q_{n}^{(i)}$, we calculate $\rho_{n}^{(i)}$ as (note that the top node connects to $k$ children since it has no parent, and its degree is distributed according to $p_{k}^{(i)}$ instead of $\tilde{p}_{k}^{(i)}$)

\begin{equation}
\begin{gathered}
    \rho_{n}^{(i)} = \rho_0^{(i)} + (1-\rho_0^{(i)}) \sum_{k}p_{k}^{(i)}\sum_{m=\ceil*{\theta k}}^{k}\binom{k}{m}(\bar{q}_{n}^{(i)})^m \\
    \times (1-\bar{q}_{n}^{(i)})^{k-m} \equiv h^{(i)}(\bar{q}_{n}^{(i)}).
\end{gathered}
\end{equation}

In synchronous updating ($f=1$), the diffusion speed in $C_i$ at time $t$ can be approximated as: $v_t^{(i)}=d\rho_{t}^{(i)}/dt = [\rho_{t+1}^{(i)}-\rho_{t}^{(i)}]^{+}$, where the notation $[\cdot]^{+}$ stands for $\max(0, \cdot)$. 
The overall diffusion speed $v_t$ at time $t$, the total diffusion time $t_s$, and the average diffusion speed $\bar{v}$ are

\begin{align}
    v_t = \sum_{i}\frac{|C_{i}|}{N}v_t^{(i)}, \qquad
    t_s = t \mid v_t = 0, \qquad
    \bar{v} = \frac{\rho_{t_{s}} - \rho_{0}}{t_s}.
\end{align}

These equations can be adapted for asynchronous updating, provided that the fraction $f$ of nodes updated at each time step is sufficiently small such that they may be considered to be independent of each other \cite{gleeson2008cascades}. 
We introduce the following notation: $\bar{q}(t)$, $q(t)$ and $\rho(t)$. The evolution equations for asynchronous updating are
\begin{align}
    \bar{q}^{(i)}(t) &= \frac{1}{d}\sum\nolimits_{j}e_{ij}q^{(j)}(t-1),\\
    dq^{(i)}(t)/dt &= f[g^{(i)}(\bar{q}^{(i)}(t+1)) - q^{(i)}(t)]^{+},\\
    v^{(i)}(t) = d\rho^{(i)}(t)/dt &= f[h^{(i)}(\bar{q}^{(i)}(t+1)) - \rho^{(i)}(t)]^{+}, \label{step-speed}
\end{align}
with $q^{(i)}(0) = \rho^{(i)}(0) = \rho_{0}^{(i)}$. The speed is calculated as,

\begin{align} \label{avg-speed}
    v(t) = \sum_{i}\frac{|C_{i}|}{N}v^{(i)}(t),\qquad
    \bar{v} = \frac{\rho({t_{s}}) - \rho(0)}{t_s}.
\end{align}

\newpage

\section{Results}

\subsection{Network size}

\begin{figure*}[ht!]
\centering
\includegraphics[trim=0mm 0mm 0mm 0mm, width=0.8\columnwidth]{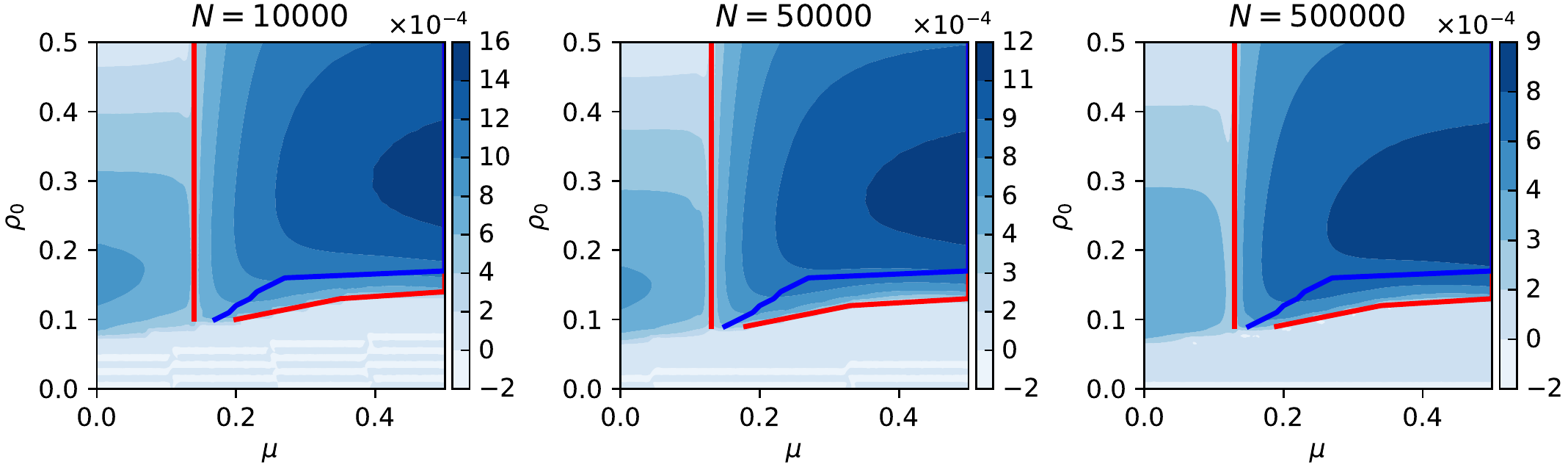}
\caption{Phase diagrams of the average diffusion speed on SBM networks with different number of nodes $N$. The results are derived from the analytical approach. 
Other model parameters are: $z=10, \theta=0.35, f=0.01$.}
\label{fig-si-3}
\end{figure*}

\begin{figure*}[ht!]
\centering
\includegraphics[trim=0mm 0mm 0mm 0mm, width=0.8\columnwidth]{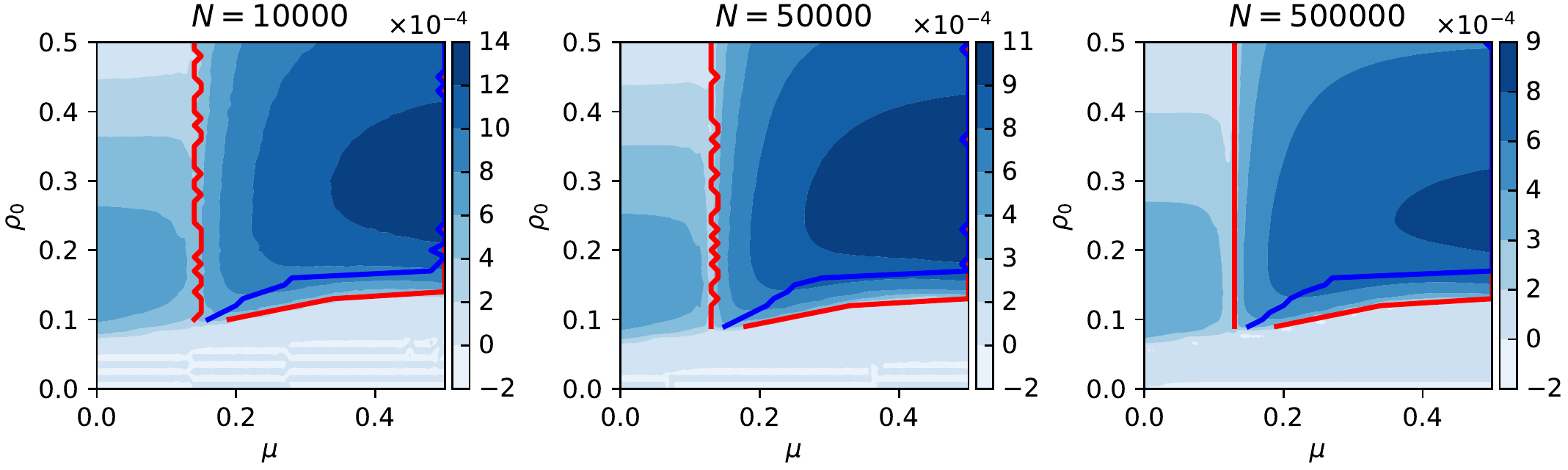}
\caption{Phase diagrams of the average diffusion speed on SBM networks of different sizes $N$, derived from numerical simulations (averaged over $100$ runs). 
Other model parameters are: $z=10, \theta=0.35, f=0.01$.}
\label{fig-si-4}
\end{figure*}

Figure~\ref{fig-si-3} and Figure~\ref{fig-si-4} present results based on SBM networks with different number of nodes, derived through the analytical approach and the numerical simulation, respectively.
It shows that the network size does not change our finding of the most efficient spreading behavior with respect to the network modularity.

\subsection{Average degree}

\begin{figure*}[ht!]
\centering
\includegraphics[trim=0mm 0mm 0mm 0mm, width=0.8\columnwidth]{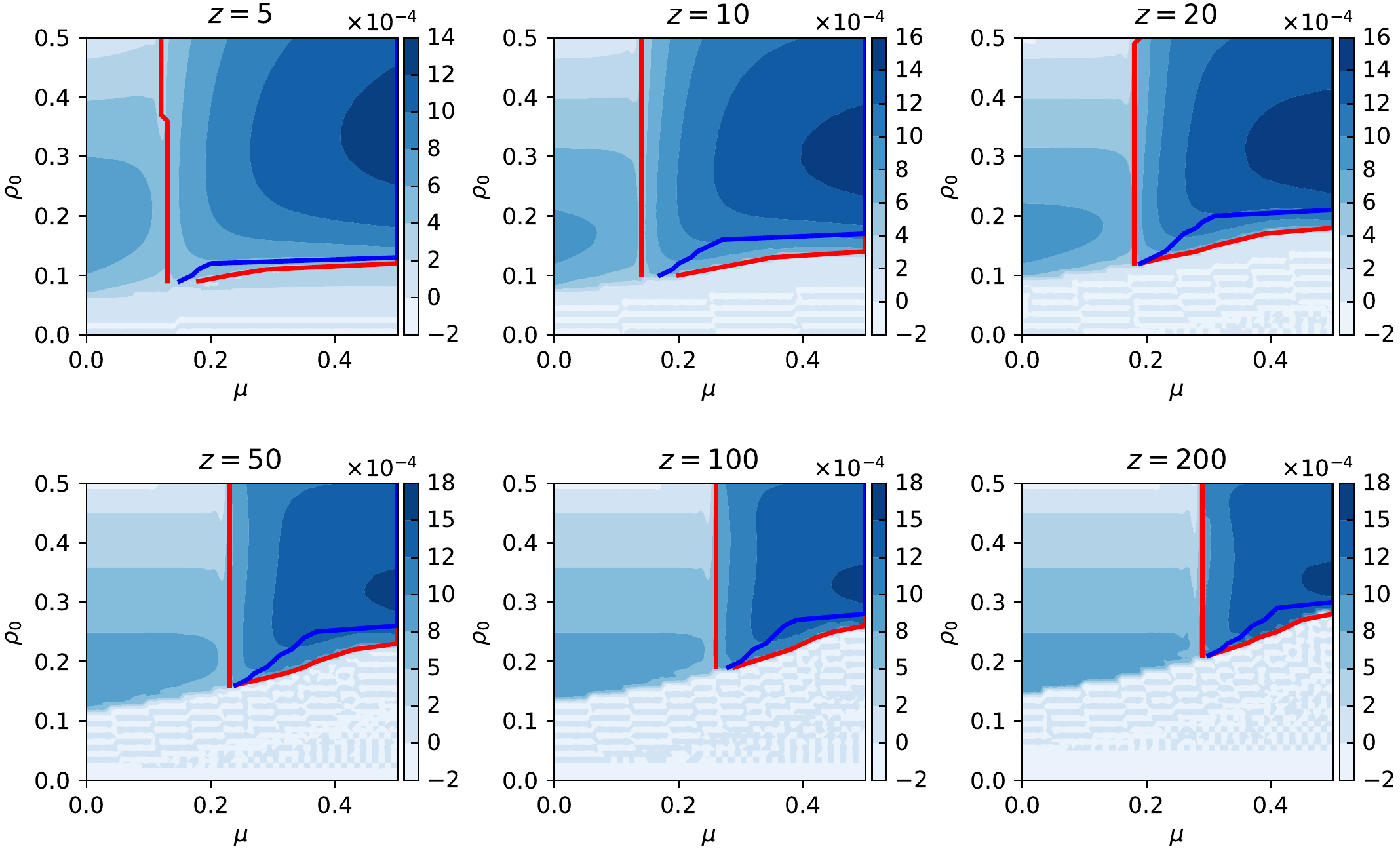}
\caption{Phase diagrams of the average diffusion speed on SBM networks derived from the analytical approximation. 
Each subplot corresponds to networks with a specific average degree $z$.
Other model parameters are: $N=1\times10^4, \theta=0.35, f=0.01$.}
\label{fig-si-1}
\end{figure*}

\begin{figure*}[ht!]
\centering
\includegraphics[trim=0mm 0mm 0mm 0mm, width=0.3\columnwidth]{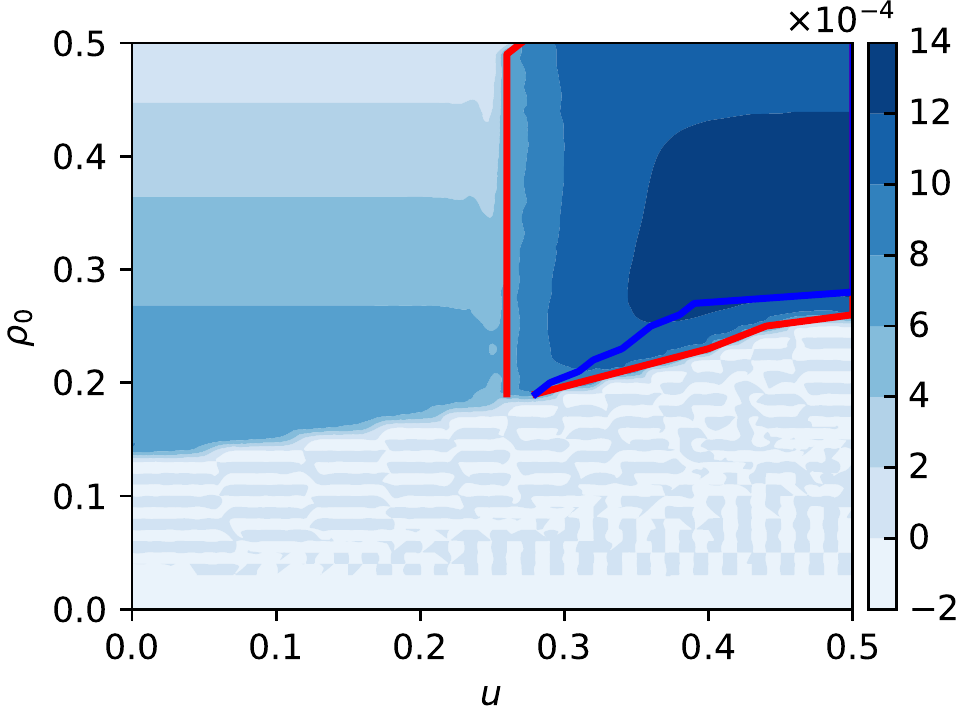}
\caption{Phase diagrams of the average speed of information diffusion on SBM networks with average degree $z=100$.
The results are obtained from numerical simulations. 
Other model parameters are: $N=1\times10^4, \theta=0.35, f=0.01$.}
\label{fig-si-2}
\end{figure*}

Figure~\ref{fig-si-1}-\ref{fig-si-2} show the average diffusion speed as a function of the seed size and the network modularity, on SBM networks with different average degrees.
The results indicate that, as one increases the average degree, the minimal number of inter-community edges (or the maximum modularity) required to generate global cascades also increases, so does the minimal number of seeds.
This is expected because more active neighbors are needed to achieve the same adoption threshold when the nodes' neighbor size increases.

However, the optimal network modularity for the overall fastest information diffusion always exists when global cascades are enabled.
And the optimal value depends on the seed size, which agrees with our finding in the main text.
In other words, the average degree does not change the behavior of our system qualitatively. 

\subsection{Anti-modular SBM networks}

\begin{figure*}[ht!]
\centering
\includegraphics[trim=0mm 0mm 0mm 0mm, width=0.6\columnwidth]{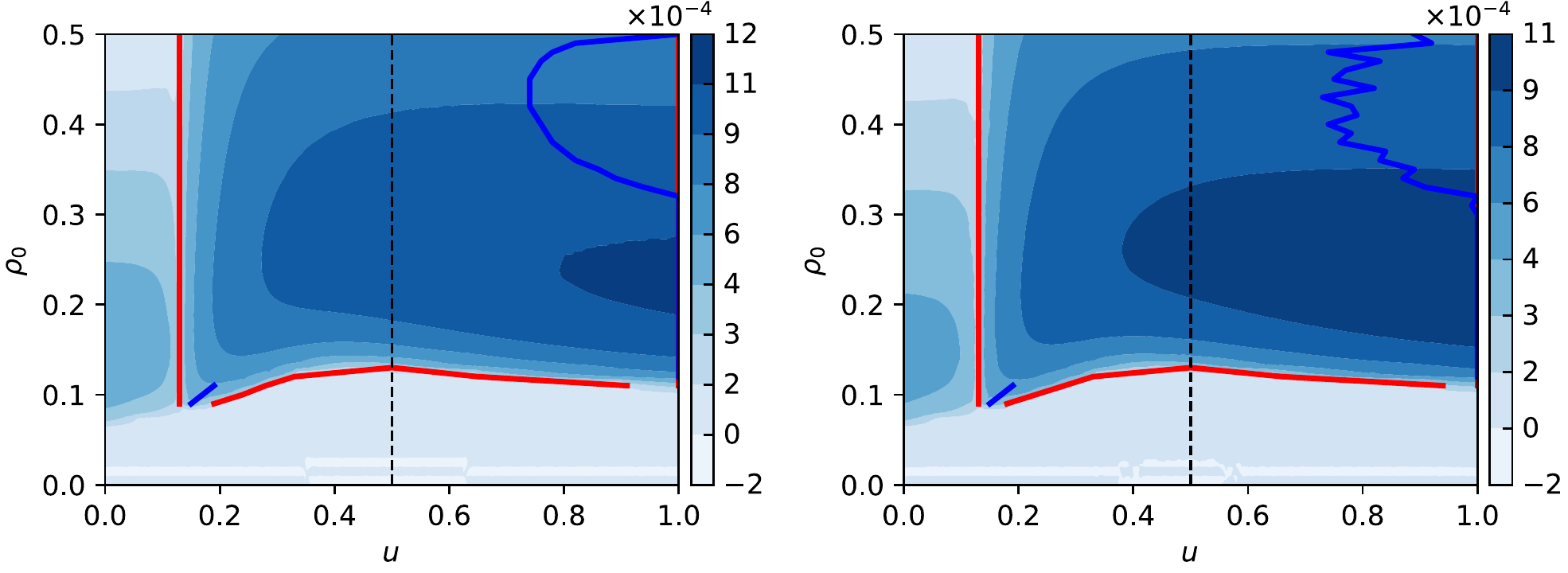}
\caption{Phase diagrams of the average diffusion speed in the whole range of $\mu$ in SBM networks. The results are based on both analytical predictions (left) and numerical simulations (right), averaged over $100$ runs.
Model parameters are: $N=1\times10^5, z=10, \theta=0.35, f=0.01$. 
There are three regions: $\mu<0.5$ (assortative and modular); $\mu=0.5$ (random); and $\mu>0.5$ (disassortative and anti-modular).
The anti-modular networks behave quite differently from the modular networks.}
\label{fig-si-bi}
\end{figure*}

Although our focus is on networks with community structure, it is intriguing to examine the diffusion dynamics on anti-modular networks.
Figure~\ref{fig-si-bi} shows the average diffusion speed in the whole range of $\mu$ on SBM networks, where the network shifts from exhibiting a modular structure to displaying a bipartite structure.
Interesting patterns emerge: different from the dynamics on modular networks, where information spreads from the originating community to the other, the diffusion process on anti-modular networks temporally alternates between the two communities. 

In such a scenario, global cascades still require a minimal number of seeds, but unlike modular networks, when $\rho_0$ is not too large (e.g., $\rho_0=0.2$), strong anti-modular structure (large $\mu$) always promotes the diffusion speed, making the strict bipartite networks the ideal conditions for global cascades. However, when $\rho_0$ is sufficiently large (e.g., $\rho_0=0.4$), the most efficient global cascade happens at an intermediate strength of anti-modular structure (Figure~\ref{fig-si-bi}). 

\subsection{Seed arrangement}

\begin{figure*}[ht!]
\centering
\includegraphics[trim=0mm 0mm 0mm 0mm, width=0.3\columnwidth]{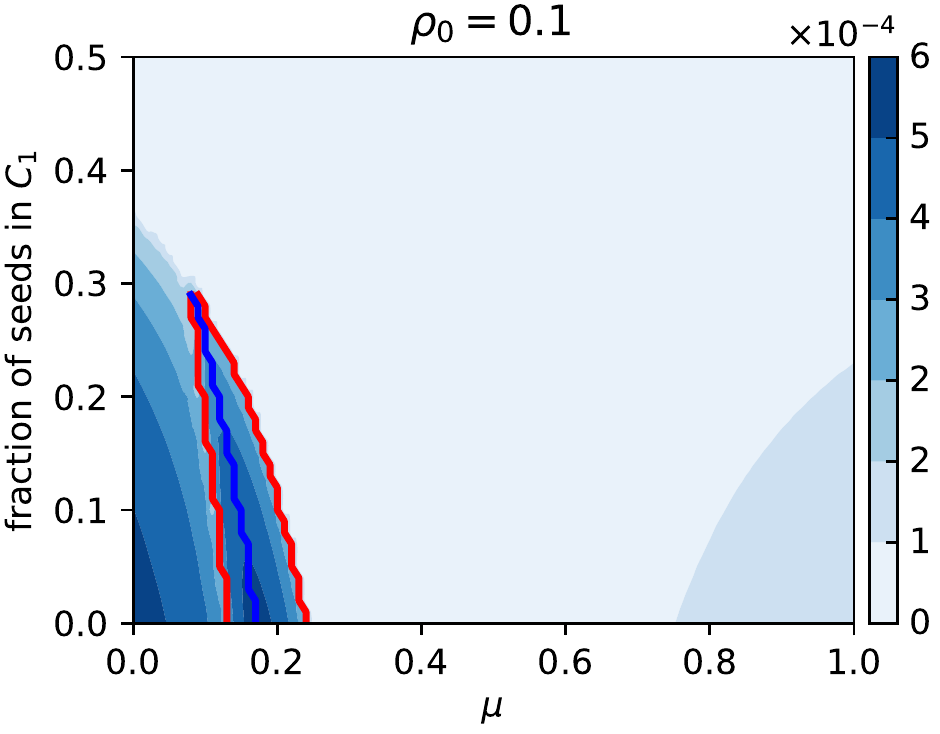}
\caption{Phase diagrams of the average information diffusion speed in the whole range of $\mu$, as a function of seed arrangements between two communities in SBM networks. 
The results are based on analytical predictions. 
The $y$-axis represents the fraction of seeds placed in $C_1$. 
The seed size $\rho_0 = 0.1$.
Other model parameters are: $N=1\times10^5, z=10, \theta=0.35, f=0.01$}
\label{fig-si-seed-arr}
\end{figure*}

We also examine the diffusion dynamics in our system under conditions where the seeds are not entirely placed in a single community.
Figure~\ref{fig-si-seed-arr} shows that, at any given seed distribution in the network (draw a horizontal slice), when global cascades are possible, there is a window of network modularity for information diffusion at global scale. 
For example, when all seeds are placed in $C_2$ (none in $C_1$), the $\mu$ window for global cascades is $[0.13, 0.24]$, and the fastest diffusion process happens at a middle level of modularity ($\mu = 0.17$), which is exactly what we see in Fig.~1 in the main text.
The same pattern holds for other seed arrangements in Figure~\ref{fig-si-seed-arr}.
In other words, our finding of an intermediate strength of network modularity being the ideal condition for efficient global cascades can be generalized to all other seed arrangements in the two communities, for the seed size $\rho_0 = 0.1$.

\subsection{Number of communities}

So far, all our experiments on SBM networks are limited to the case of two equally sized communities ($|C_1|=|C_2|$).
Here, we examine the diffusion dynamics on SBM networks with different number of communities.
As a first step, we assume that all communities still have the same number of nodes and links are randomly placed according to the parameter $\mu$, as is the case in the main text.
The mixing matrix is:
\begin{equation}
\mathbf{e} = \frac{1}{d}
\begin{bmatrix} 
1-\mu  & \frac{\mu}{d-1} & \dots \\
\vdots & \ddots    &       \\
\frac{\mu}{d-1}  &      & 1-\mu
\end{bmatrix},
\end{equation}
where $\mathbf{e}$ is $d \times d$ and $d$ is the number of communities. 
The diagonal entries of $\mathbf{e}$ are $\frac{1-\mu}{d}$ and the off-diagonal entries are $\frac{\mu}{d(d-1)}$~\cite{newman2003mixing}. 
The network modularity can be calculated as: $Q=1-\mu-\frac{1}{d}$, which means that, in order to generate modular networks, $\mu$ can be larger than $\frac{1}{2}$ when $d$ is large than $2$.

\begin{figure*}[ht!]
\centering
\includegraphics[trim=0mm 0mm 0mm 0mm, width=0.8\columnwidth]{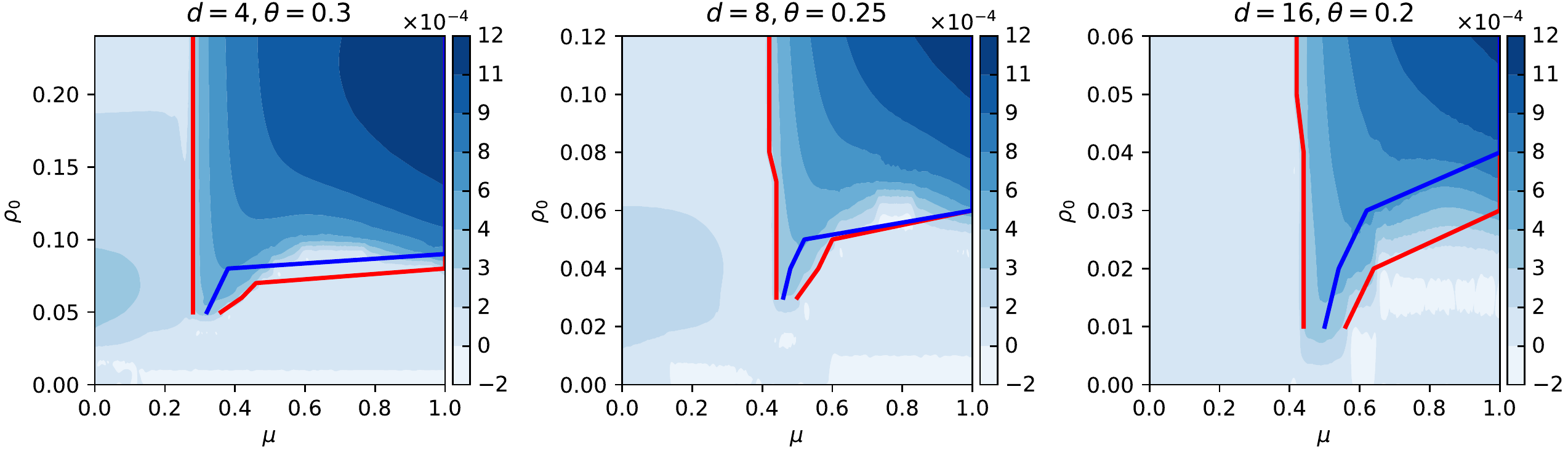}
\caption{Phase diagrams of the average diffusion speed on SBM networks with different number of equally sized communities $d$, derived from the analytical approximation. Seeds are randomly selected from a single community. Other model parameters are: $N=1\times10^5, z=10, f=0.01$.}
\label{fig-si-5}
\end{figure*}

Figure~\ref{fig-si-5} shows the analytical results of the average diffusion speed on SBM networks with different number of communities.
Please note that, at any given $\mu$, the number of bridges running between a pair of communities decreases as the number of communities $d$ increases.
Thus networks with more communities require smaller adoption threshold $\theta$ in order to achieve global cascades.
Figure~\ref{fig-si-5} indicates that our finding of the optimal network modularity for the most efficient global diffusion can generalize to networks with multiple communities.

\subsection{Simulations on real-world networks}

In the main paper, we showed simulation results on LFR and Twitter networks. Here we extend our experiments to more real-world networks across different domains including social, communication, and collaboration networks from~\cite{snapnets}.
Like the case of Twitter, we use the largest connected component (LCC) of the undirected version of each network, and use the parameter $p$ to control the network modularity through the edge rewiring process described in the main text.
Note that we focused on networks with a LCC that contains at least $50$K nodes and excluded those with more than $1$M nodes to make the simulations feasible and comparable to SBM and LFR networks.
Table~\ref{tab:net-stat} summarizes the statistics of networks we test here.

\begin{table}[]
\begin{tabular}{|l|r|r|r|r|r|}
\hline
\multicolumn{1}{|c|}{\textbf{Network Name}} & \multicolumn{1}{c|}{\textbf{Num. of Nodes}}   & \multicolumn{1}{c|}{\textbf{Num. of Edges}}    & \multicolumn{1}{c|}{\textbf{Avg. Degree}}  & \multicolumn{1}{c|}{\textbf{Num. of Communities}} & \multicolumn{1}{c|}{$Q_{\textbf{norm}}$} \\ \hline
DBLP                       & 317,080                      & 1,049,866 &  6.6 & 239      & 0.84 \\ \hline
Eu Email                   & 224,832                      & 339,925  & 3  & 89        & 0.80 \\ \hline

Slashdot                   & 82,168                       & 504,230  &  12.3 &  549       & 0.44                            \\ \hline
Twitter                    & 81,306                       & 1,342,310  & 33.0 & 70   & 0.86                            \\ \hline
Epinions                   & 75,877                       & 405,739 & 10.7  & 776  & 0.55                            \\ \hline
Deezer                     & 54,573                       & 498,202  & 18.3  & 24   & 0.79                            \\ \hline
FB Pages                   & 50,515                       & 819,090  & 32.4  & 34     & 0.72                            \\ \hline
\end{tabular}
\caption{Statistics of the largest connected component of seven real-world networks we tested. Directed networks are all converted to undirected networks. The communities are detected using the Louvain algorithm~\cite{blondel2008fast}. The community sizes are heterogeneous. Note that the Twitter network has been used in the main paper.}
\label{tab:net-stat}
\end{table}

\begin{figure*}[ht!]
\centering
\includegraphics[trim=0mm 0mm 0mm 0mm, width=0.8\columnwidth]{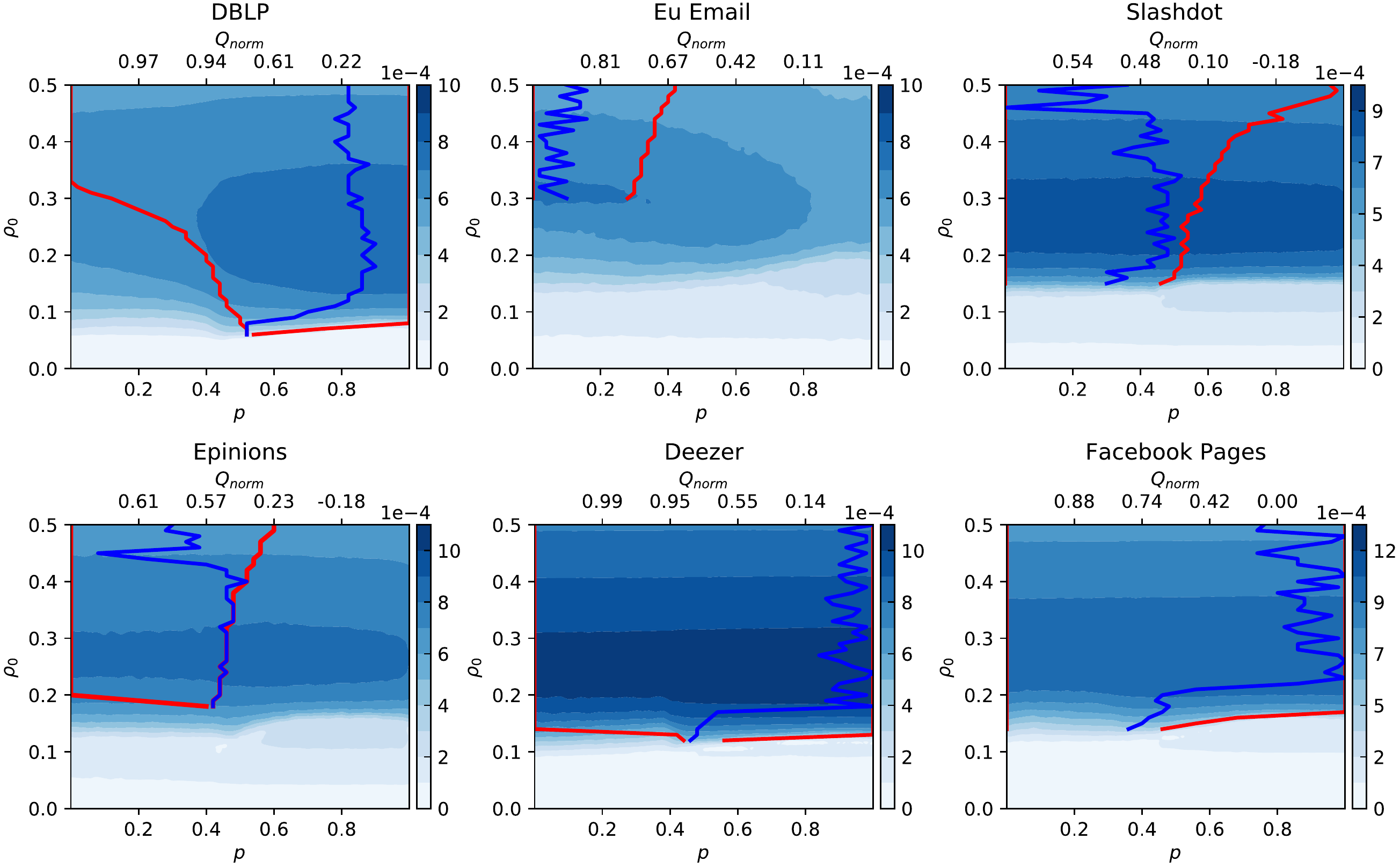}
\caption{Phase diagrams of the average diffusion speed $\bar{v}$ on six real-world networks. Network statistics are shown in Table~\ref{tab:net-stat}. 
Network modularity is controlled by parameter $p$ on the $x$-axis, with the corresponding normalized modularity $Q_{\text{norm}}$ shown on the top axis. 
The blue curve indicates the optimal $p$ for $\bar{v}$ for a given seed size $\rho_0$ (there is only a single $p$ that maximizes $\bar{v}$ for any given $\rho_0$).
Simulation parameters are: $\theta=0.3, f=0.01$. Seed nodes are randomly selected across the whole network.}
\label{fig-si-real}
\end{figure*}

Figure~\ref{fig-si-real} shows the phase diagrams for six empirical networks. 
The pattern looks similar to that on SBM, LFR, and Twitter networks. There exists an optimal modularity for overall fast global cascades, and the optimal value depends on the seed size and the network.


\subsection{Average diffusion speed with a constraint on time}

There are many real world diffusion applications that need to be optimized for the speed with a predefined cascade size. However, there are also cases where one cares about the speed with a time limit (or equivalently the cascade size for a fixed time window). For example, a get-out-the-vote campaign on election day may need to be optimized for adoption speed since the operation will be useless after the election is over.
Additionally, the spread of health behaviors such as wearing masks and social distancing aims to slow down the spread of Coronavirus \emph{before} hospital capacity is surpassed.
We thus examine the optimal structure for diffusion speed with a time constraint.

\begin{figure*}[ht!]
\centering
\includegraphics[trim=0mm 0mm 0mm 0mm, width=0.8\columnwidth]{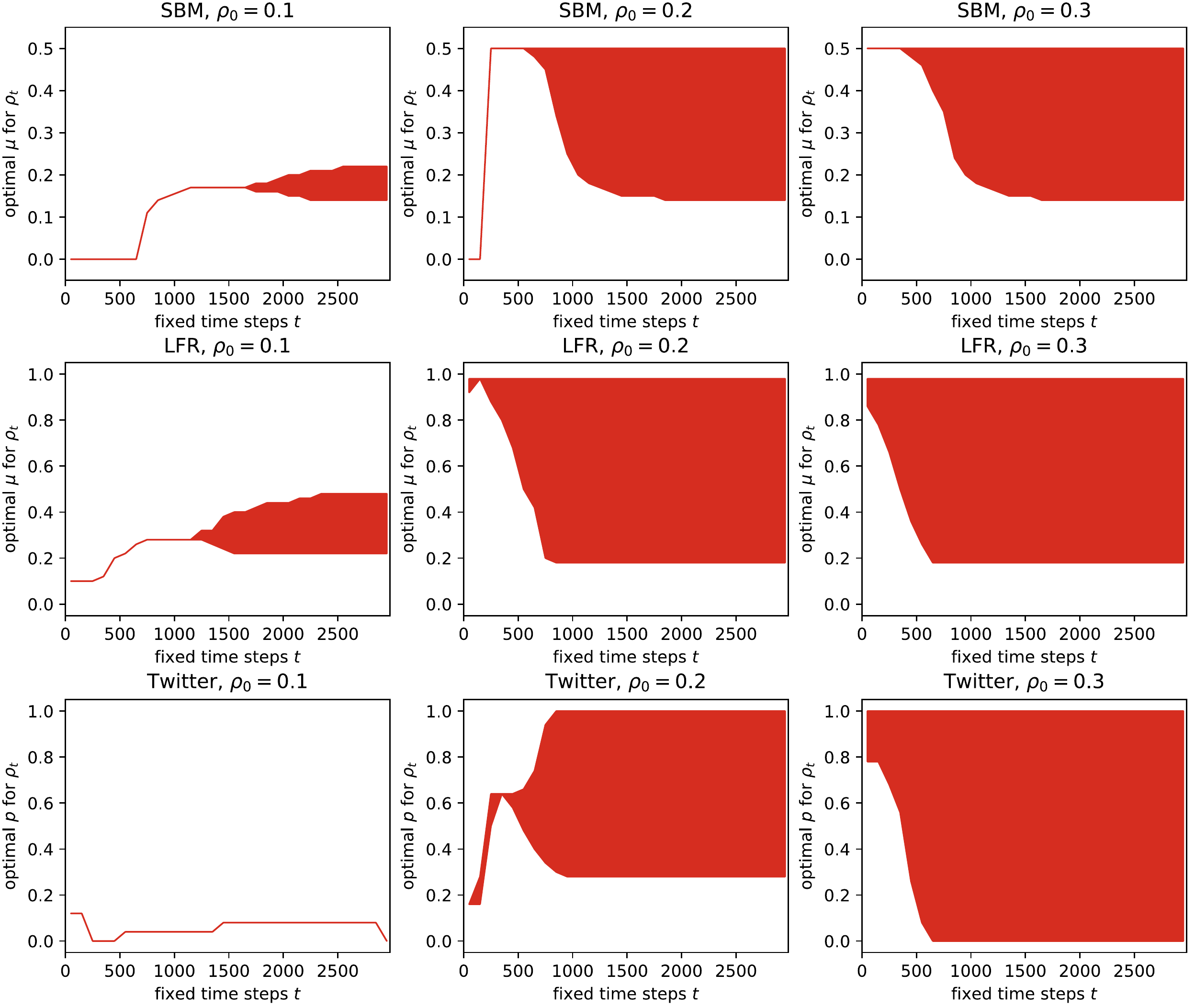}
\caption{The optimal modularity for the average diffusion speed with a constraint on diffusion time. The results are based on simulations. The most efficient network transitions from having a single optimal modularity to exhibiting a wide range of optimal modularity (controlled by $\mu$ or $p$) as the time increases, especially for small seed sizes.
This trend is qualitatively the same for synthetic networks (SBM, LFR) and real-world networks (Twitter).}
\label{fig-si-fix-time}
\end{figure*}

Figure~\ref{fig-si-fix-time} indicates that the best modularity for fast diffusion also tends to decrease as the diffusion time increases. 
For instance, the optimal $\mu$ changes from $\mu=0$ to $\mu=0.17$ as $t$ increases from $t=500$ to $t=1500$ for $\rho_0=0.1$ on SBM networks.
In other words, if the goal is to infect as many nodes as possible in a very short period of time, then a higher modularity is better than the optimal for a longer time window when global cascades are achieved. 
The intuition behind this result is that since the diffusion speed within communities is higher than that for inter-community spreading at the early stage of the diffusion process (see Fig.~2 in the main paper), when the time available for diffusion is limited, it is better to have strong modularity to promote local spreading.

What's more, unlike conditions with a constraint on cascade size where the optimal modularity is typically a single value, when the constraint is on time, networks can exhibit a wide range of optimal modularity values, especially for a large time budget. 
Intuitively, if the time is sufficiently long, many modularity values are optimal as long as they are  within the window of global cascades. 
When the seed size is too large (e.g., $\rho_0=0.3$), there tends to exist a wide range of optimal modularity values, regardless of the time budget. The reason is that the diffusion process tends to reach global cascades so quickly that the time budget usually cannot be exhausted.

\bibliographystyle{plain}
\bibliography{supp}